\documentclass[journal,twoside]{IEEEtran}
\usepackage[T1]{fontenc}
\usepackage{cite}
\usepackage{amsmath,amssymb,amsfonts}
\usepackage{algorithmic}
\usepackage[]{graphicx}
\usepackage{epsfig}
\usepackage{TUSON}
\usepackage[utf8]{inputenc}
\usepackage{amsmath,amssymb}
\usepackage{graphicx}

\usepackage{amsmath,amssymb,amsfonts}
\usepackage{algorithmic}
\usepackage{textcomp}
\usepackage{wrapfig,colortbl}
\usepackage[switch,pagewise]{lineno}

\usepackage[caption=false,font=footnotesize]{subfig}
\usepackage{booktabs}
\usepackage{array}
\usepackage{cite}
\usepackage{url}
\usepackage{multirow}
\usepackage{xr}
\graphicspath{{figures/}}
\usepackage[export]{adjustbox}
\usepackage{soul}

\title{MRI-Free CT-to-Ultrasound Skull Registration in Transcranial Focused Ultrasound using CMUT based Integrated Therapeutic and Imaging Arrays}

\author{Reza Pakdaman Zangabad\IEEEauthorrefmark{1},~\IEEEmembership{Member,~IEEE},
Sai Vikas Satyanarayana\IEEEauthorrefmark{1},
Costas D.~Arvanitis,~\IEEEmembership{Member,~IEEE},
and F. Levent Degertekin,~\IEEEmembership{Fellow,~IEEE}%
\thanks{This work was supported by the National Institutes of Health (NIH), under Grant R37CA239039.}
\thanks{\IEEEauthorrefmark{1}Reza Pakdaman Zangabad and Sai Vikas Satyanarayana contributed equally to this work.}
\thanks{R. Pakdaman Zangabad and S. V. Satyanarayana are with the Mechanical Engineering Department, Georgia Institute of Technology, Atlanta, GA 30332 USA (e-mail: rpz3@gatech.edu; saivikas@gatech.edu).}
\thanks{C. D. Arvanitis is with the Biomedical Engineering and Mechanical Engineering Departments, Georgia Institute of Technology, Atlanta, GA 30332 USA (e-mail: costas.arvanitis@gatech.edu).}
\thanks{F. L. Degertekin is with the Mechanical and Electrical and Computer Engineering Departments, Georgia Institute of Technology, Atlanta, GA 30332 USA (e-mail: levent.degertekin@me.gatech.edu).}
\thanks{Corresponding author: Reza Pakdaman Zangabad (e-mail: rpz3@gatech.edu).}
}

\IEEEaftertitletext{\GA{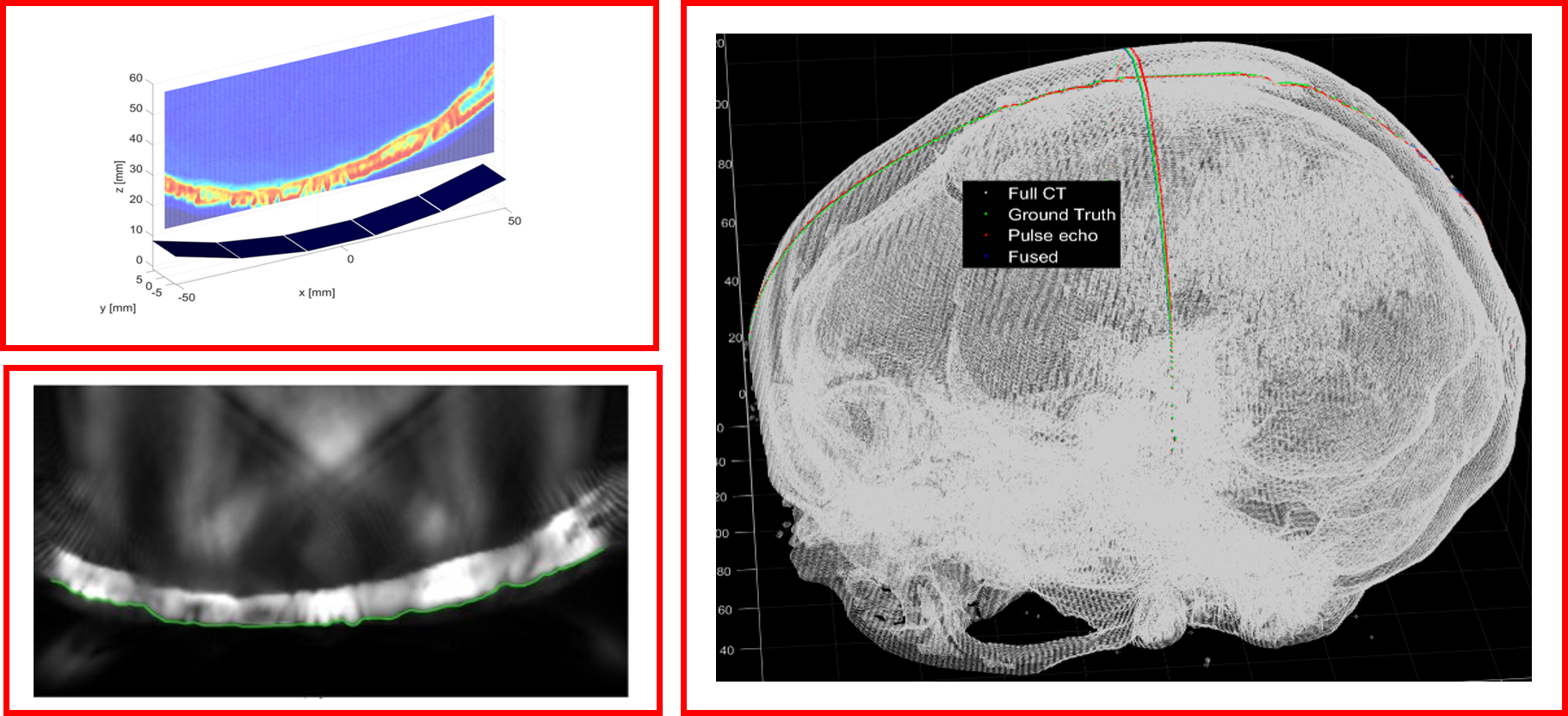}
\Abstract{\begin{abstract}
Accurate skull-to-array registration is essential for transcranial focused ultrasound (tFUS) because CT-derived skull models must be aligned with the physical-array coordinate system to support patient-specific aberration correction. Current workflows typically use MRI and/or external tracking to establish this spatial relationship, adding imaging infrastructure and coordinate transformations to the treatment workflow. Here, we describe an MRI-free CT-to-ultrasound registration framework utilizing a hybrid hemispherical CMUT tFUS array that combines a sparse therapeutic array with an integrated CMUT US-Cross imaging array. The therapeutic array provides pulse-echo measurements of skull location, while the US-Cross array provides local high-resolution images of the skull surface. The approach is evaluated on a CT-derived outer cortical skull surface registered to ultrasound-derived features across a 25-pose matrix with 10-mm lateral offsets between the skull and array centers and ±5◦ angular perturbations. Five randomized radial-noise seeds were applied to therapeutic array pulse-echo measurements to model measurement or element-placement errors, yielding 125 pose-seed realizations per noise and active-element setting. US-Cross imaging features were generated using Field II and delay-multiply-and-sum plane-wave compounding. Four registration modes were evaluated: Therapeutic only, US-imaging only, Sequential, and Fused registration. Fused registration, combining both therapeutic and US-imaging information, yielded the best overall performance, with a median total displacement error of 0.271 mm [0.169--0.420] and a median rotation error of 0.904$^{\circ}$ [0.432--1.639]. These results support an all-ultrasound-based approach using a hybrid CMUT tFUS array for MRI-free CT-US skull registration.
\hblne
\end{abstract}}
\vspace{1\baselineskip}\vspace*{-1pt}
}

\begin{document}

\maketitle

\Highlights[1]{An MRI-free CT-to-ultrasound skull registration framework was developed for a hybrid hemispherical CMUT tFUS array integrating sparse therapeutic pulse-echo sensing with high-frequency CMUT imaging.}

\Highlights[2]{ Fused registration reduced median total displacement error from 0.518 mm [0.278--0.961] to 0.271 mm [0.169--0.420] and median rotation error from 1.440$^\circ$ [0.749--2.505] to 0.904$^\circ$ [0.432--1.639]. }

\Highlights[3]{Integrated high-frequency US-Cross imaging provided a stabilizing local geometric constraint that improved CT-US skull registration robustness across pose, radial-noise, and active-element conditions.}

\Keywords{Aberration correction, capacitive micromachined ultrasonic transducers (CMUTs), CT-to-ultrasound registration, focused ultrasound, image-guided therapy, iterative closest point (ICP), skull registration, transcranial focused ultrasound (tFUS), ultrasound imaging.}

\PrintHighlights

\section{Introduction}

Transcranial focused ultrasound (tFUS) enables non-invasive delivery of acoustic energy to localized brain targets and has been investigated for thermal ablation, neuromodulation, blood-brain barrier opening, sonobiopsy, and histotripsy \cite{elias2016randomized,legon2014somatosensory,hynynen2001noninvasive,pacia2022sonobiopsy,sukovich2016histotripsy}. Across these applications, skull geometry and composition distort acoustic propagation through refraction, reflection, attenuation, mode conversion, and phase delay; if uncorrected, these effects can shift or broaden the focus, reduce peak pressure, and increase unintended energy deposition \cite{hynynen1998potential,clement2002noninvasive,aubry2003experimental}. These effects can be accounted for by employing phased-array systems and patient-specific CT to segment the skull, estimate acoustic properties, and calculate element-wise phase and amplitude corrections using ray-based, pseudo-spectral, finite-difference, or hybrid acoustic models \cite{schoen2022towards,aubry2003experimental,pichardo2011multifrequency,jones2016comparison,bancel2021comparison,sakharova2025alendronate}. Accurate skull-to-array registration makes these CT-derived corrections physically meaningful in the treatment frame.

\begin{figure}[!t]
\centering
\includegraphics[width=0.47\textwidth]{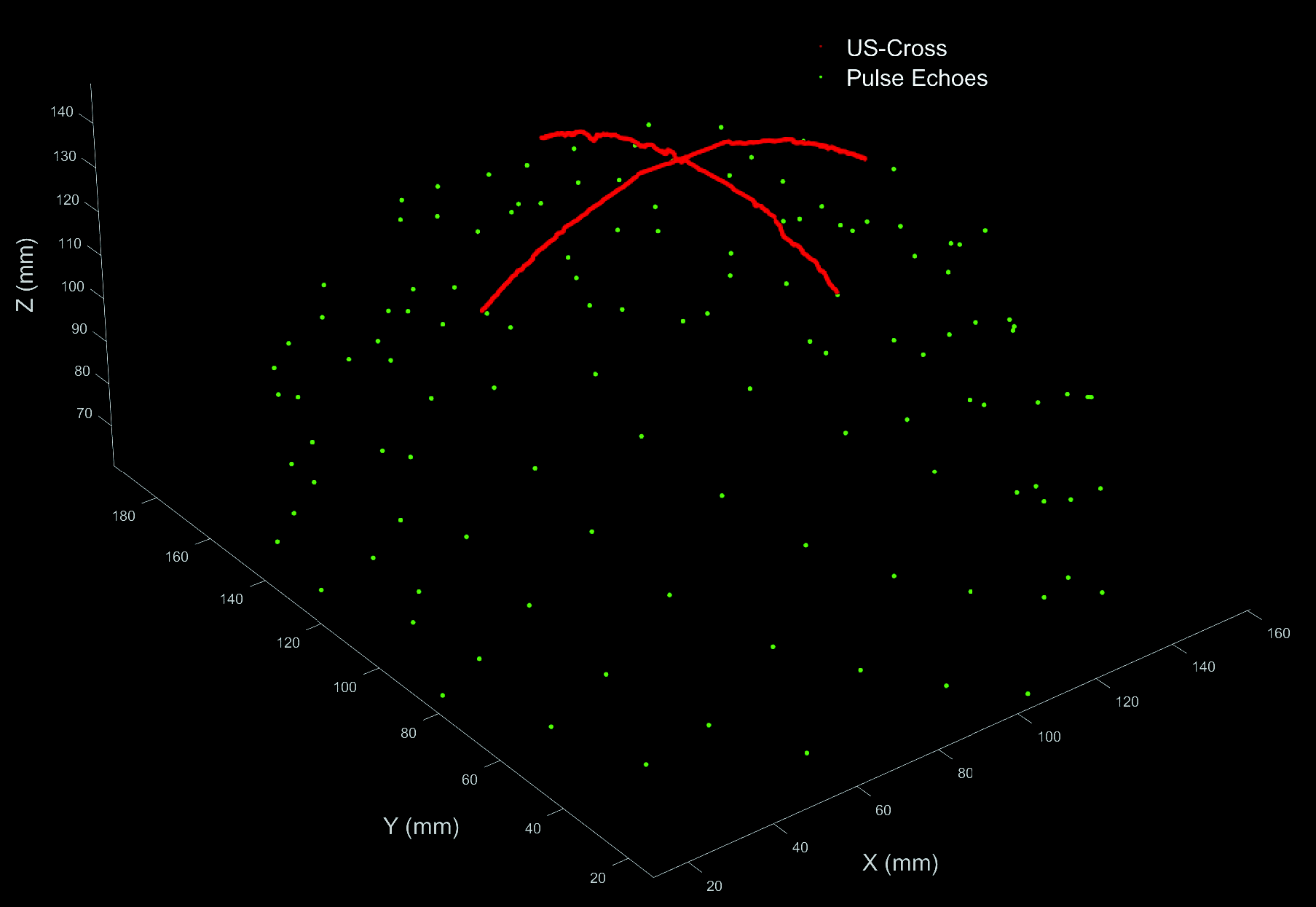}
\caption{System concept and sensor geometry: fixed hemispherical tFUS array, sparse therapeutic elements, and nine integrated CMUT imaging arrays.}
\label{fig:system}
\end{figure}

Current workflows typically align preoperative CT with the treatment geometry using MRI with stereotactic immobilization, or frameless optical tracking and neuronavigation \cite{chaplin2019accuracy,xu2023targeting,roberts1986frameless,wu2018efficient}. In MR-guided focused ultrasound, CT provides skull morphology and acoustic-property information, whereas MRI provides target definition and the spatial reference needed to localize the target relative to the phased array; consequently, CT-to-MR registration and reproducible head-to-array positioning are required \cite{arvanitis2013combined,xu2023targeting}. Frameless optical approaches improve flexibility but introduce additional transformations among image, head, tracker, transducer, and acoustic-focus coordinate systems. Their limitations are therefore best considered as part of the full targeting chain: stereotactic frames add setup time, discomfort, and procedural rigidity; optical tracking requires line-of-sight and is sensitive to tracker calibration, fiducial geometry, image-to-patient registration, and marker or skin motion; and MRI-based workflows require dedicated imaging infrastructure and remain subject to MR geometric distortion and CT-to-MR coregistration uncertainty \cite{west2001fiducial,chaplin2019accuracy,putz2024quality}. As a benchmark from high-precision cranial stereotactic imaging, high-resolution MRI-to-CT registration has been reported at $\leq$0.5~mm in-plane and $\leq$1~mm through-plane under optimized conditions, while errors on the order of 2~mm are reported more generally \cite{putz2024quality}. By comparison, end-to-end optically guided FUS targeting errors of approximately 2.4--3.3~mm have been measured in phantom experiments, illustrating how calibration and registration errors can compound even when individual tracking components have submillimeter precision \cite{chaplin2019accuracy}. For repeated or outpatient interventions, the motivation for an integrated acoustic registration method is therefore not simply lower hardware cost, but a simpler and more reproducible workflow that establishes the skull pose directly in the treatment coordinate system with fewer external transformations.

A direct acoustic alternative is to acquire ultrasound measurements in the treatment coordinate system and register the CT-derived skull surface to the array \cite{arvanitis2013combined}. O'Reilly et al. demonstrated this approach with pulse-echo measurements from five ex vivo human skullcaps using a sparse 128-element, 11-MHz hemispherical localization array, achieving 0.9 $\pm$ 0.2~mm displacement error and 0.8 $\pm$ 0.4$^{\circ}$ rotation error relative to landmark-based registration \cite{oreilly2016registration,besl1992method}. For a 230-kHz therapeutic system, those registration errors were predicted to produce a targeting error of 1.0 $\pm$ 0.2~mm with only modest focal-pressure loss \cite{oreilly2016registration}. The full registration experiment used ex vivo skullcaps without overlying scalp under controlled degassed-water coupling, so these values represent a controlled geometric benchmark rather than clinical end-to-end performance. O'Reilly et al. further noted that an approximately 1-mm targeting error corresponded to about 26\% of the lateral $-6$-dB focal width at 230~kHz but about 74\% at 650~kHz, illustrating why the adequacy of a registration error is application- and frequency-dependent \cite{oreilly2016registration}. Crake et al. extended the concept with a dual-mode sparse hemispherical array for active 5-MHz skull localization and passive cavitation mapping, reporting skull localization within approximately $\pm$2~mm and 2$^{\circ}$ while also highlighting the rotational ambiguity of sparse measurements on an approximately hemispherical skull \cite{crake2018dualmode}. Deng et al. further integrated a 128-element, 11-MHz skull-localization array into a 256-module multi-frequency therapy array, achieving 1.4 $\pm$ 0.4~mm displacement error, 2.1 $\pm$ 0.2$^{\circ}$ rotation error, and transmit targeting errors of approximately 0.9~mm at both 306 and 612~kHz \cite{deng2016multifrequency,deng2022ultrasoundguided}. Whether a given registration error is sufficient therefore depends on the therapy frequency, focal dimensions, target size, and biological endpoint rather than on a universal geometric threshold. These studies nevertheless establish ultrasound-based CT registration as a practical route toward MRI-reduced guidance and motivate additional geometric constraints to improve rotational robustness.

\begin{figure*}[!ht]
\centering
\subfloat[]{\includegraphics[height=0.39\linewidth]{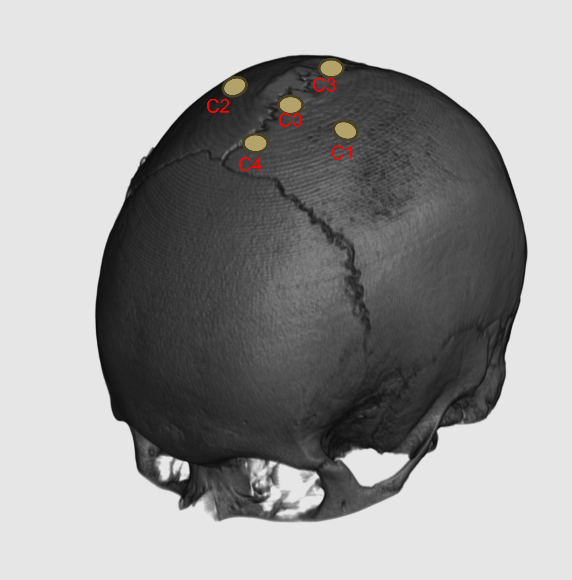}}\hfill
\subfloat[]{\includegraphics[width=0.47\linewidth]{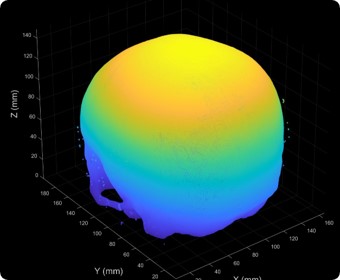}}
\caption{CT outer-surface extraction and controlled pose definition: extracted outer cortical point cloud and five skull center locations in the fixed array coordinate system.}
\label{fig:ct_pose}
\end{figure*}

Registration error is only one contributor to end-to-end targeting accuracy after skull-induced aberration correction. Lu et al. used acoustic pulse-echo co-registration within a two-step transcranial histotripsy correction workflow and reduced focal shift from 0.84 to 0.30~mm and focal volume from 10.6 to 2.0~mm$^3$ relative to no correction \cite{lu2022twostep}. Choi et al. similarly separated co-registration error from residual skull-induced focal shift in a neuronavigation-guided histotripsy system, reporting co-registration total registration error (TRE) below 2~mm, residual focal shift of approximately 0.3~mm after correction, and total targeting error of approximately 2.5~mm \cite{choi2024navith}. In a clinical TcMRgFUS thalamotomy study, Jones et al. showed that microbubble-based echo-focusing improved sonication efficiency and reduced focal size while maintaining targeting accuracy relative to conventional CT-based aberration correction \cite{jones2020echofocusing}. MRI-based end-to-end measurements provide an additional benchmark: in fresh human cadaver heads treated with a 1-MHz MR-guided HIFU system, Chauvet et al. reported mean focal-position offsets of 0.4, 0.7, and 0.5~mm along the right--left, dorsal--ventral, and rostral--caudal axes, respectively, with axis-dependent standard deviations up to 2.4~mm \cite{chauvet2013targeting}. Taken together, these studies show that geometric registration should be judged within the complete targeting chain, which also includes residual phase aberration, beam-model error, acoustic-focus localization, and treatment verification. Broader planning and monitoring studies reach the same conclusion: registration, CT-derived simulations, MR-ARFI, MR thermometry, optical tracking, patient-specific fixtures, passive acoustic mapping, cavitation imaging, and aberration-corrected transcranial imaging all depend on a shared and accurate spatial model for transmit correction, receive correction, treatment planning, and acoustic-source localization \cite{arvanitis2014transcranial,xu2023targeting,arvanitis2013combined,sigona2024planning,jiang2024scout,crake2015passive,jones2013transcranial,jiang2023ray}.

Most ultrasound-based skull registration methods still rely primarily on sparse global pulse-echo constraints from hemispherical arrays. These measurements can estimate overall skull pose, but they remain vulnerable to rotational ambiguity because the skull is smooth and approximately hemispherical, and the available echoes are sparse, angle-dependent, and affected by coupling, incidence angle, and channel-specific responses. Local conventional ultrasound imaging can provide denser geometric information, but typically over a limited field of view. To address this gap, we propose a hybrid hemispherical CMUT array for MRI-free CT-US skull registration that combines a sparse 128-element tFUS array for global pulse-echo localization with a US-Cross array layer at the center of the hemispherical frame. CMUTs are particularly suitable for this purpose, as they are broadband and the frequency response of the same CMUT array can be optimized by DC bias to be used both as therapeutic transmit array, as well as the imaging array \cite{kilinc2024cmut, kilinc2026cmut, cornu2024cmut, zangabad2024high}. Pulse-echo measurements from the therapeutic elements provide global first-return constraints, while B-mode images from the imaging arrays generate localized skull profiles that are converted into contour features for standalone, sequential, or fused registration. This work therefore evaluates whether combining sparse global pulse-echo measurements with localized high-resolution CMUT imaging improves rigid CT-US skull registration across controlled translational and rotational skull poses and multiple randomized pulse-echo radial-noise realizations, with the long-term goal of supporting compact tFUS guidance for aberration correction, targeting, cavitation monitoring, and outpatient therapy with reduced reliance on MRI. \cite{oreilly2016registration,crake2018dualmode,deng2022ultrasoundguided,lu2022twostep,choi2024navith,kilinc2026cmut,oralkan2002cmut,khuriyakub2011cmut}.

\section{Materials and Methods}

\subsection{Study design and scope}

This study evaluated an all-ultrasound-based CT-to-ultrasound skull registration workflow for a hybrid hemispherical tFUS array using a controlled pose-dependent simulation on a skull anatomy. The coordinate system of the array was fixed, and the skull CT volume was repositioned relative to the array to quantify registration behavior under translational and angular perturbations. Four registration modes were compared using identical pose inputs: Therapeutic only, US-imaging only, Sequential refinement, and Fused registration. Therapeutic only used sparse global first-return points from the 128-element therapeutic array, US-imaging only used localized high-frequency skull-edge profiles from the imaging arrays, Sequential refinement used pulse-echo registration followed by US-Cross refinement, and Fused registration optimized both feature sets jointly.

\subsection{System geometry and coordinate definition}

The modeled system consisted of a 15-cm-radius hemispherical array containing a sparse 128-element therapeutic tFUS array and an integrated US-Cross imaging layer (Fig.~\ref{fig:system}). The therapeutic elements provided global pulse-echo skull localization, while the US-Cross consisted of nine 3.5-MHz, 64-element CMUT imaging subarrays arranged near the center of the hemispherical frame to generate orthogonal XZ and YZ skull profiles \cite{kilinc2026cmut}. The nominal positions and orientations of the CMUT imaging subarrays relative to the therapeutic array were defined by the fixed device geometry and were treated as known exactly in the present simulation. Thus, inter-array calibration error was set to zero and was not independently perturbed. Throughout the study, the therapeutic and imaging arrays remained stationary, and the skull was treated as the moving object, matching the practical task of estimating skull pose in the physical treatment-array coordinate system.

\begin{figure*}[!ht]
\centering
\subfloat[]{\includegraphics[width=0.31\textwidth]{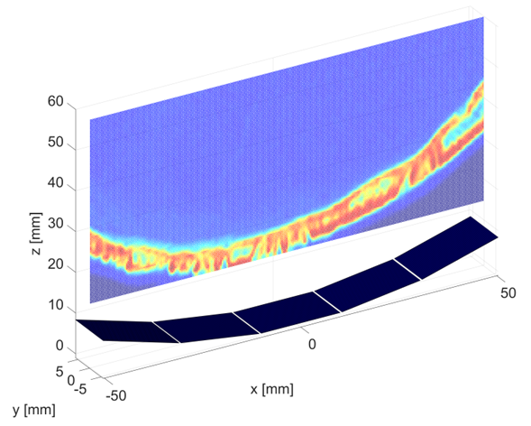}}\hfill
\subfloat[]{\includegraphics[width=0.31\textwidth]{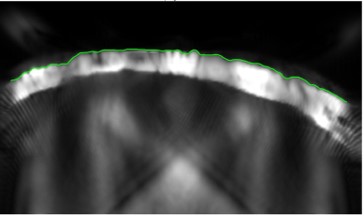}}\hfill
\subfloat[]{\includegraphics[width=0.31\textwidth]{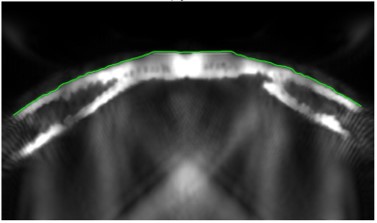}}
\caption{Representative Field II-generated B-mode images and processed skull edges in XZ and YZ planes for nominal and off-nominal poses.}
\label{fig:bmode_edges}
\end{figure*}

\subsection{CT skull model and outer-surface extraction}

A pre-segmented cranial CT volume was used to generate the anatomical target for registration. After global thresholding, the CT data were converted into a binary skull mask, and an isosurface algorithm was applied at 0.4-mm resolution to extract the skull boundary (Fig.~\ref{fig:ct_pose}). Because ultrasound registration is driven by the first strong reflection from the outer skull table, the raw CT surface was further reduced to the acoustically visible outer cortical surface. To do this, the CT point cloud was centered at the skull centroid, mapped into spherical coordinates, and discretized into high-resolution angular bins. Within each bin, candidate points were sorted by radial distance from the centroid, and only the outermost points were retained. This removed internal diploic, inner-table, and occluded structures that could otherwise bias closest-surface registration toward features not represented in the ultrasound measurements \cite{oreilly2016registration,besl1992method,deng2022ultrasoundguided}. The retained vertices were converted to physical millimeter coordinates using the CT voxel spacing and used as the fixed anatomical target for all registration modes. Surface normals were computed from the local CT surface geometry for point-to-plane registration.

\subsection{Controlled relative-pose matrix}

The primary experiment used a 25-case pose matrix formed by five skull center locations and five angular conditions. The center positions were C0, C1, C2, C3, and C4, corresponding to the nominal skull center, +X offset, -X offset, +Y offset, and -Y offset, respectively. The X and Y offsets were both 10 mm. The angular conditions were A0, A1, A2, A3, and A4, corresponding to nominal orientation, +5 degrees azimuth, -5 degrees azimuth, +5 degrees elevation, and -5 degrees elevation. This matrix was selected to separate first-order lateral positioning sensitivity from angular sensitivity while keeping the Field II simulation burden feasible. The resulting pose matrix is summarized schematically in Fig.~\ref{fig:ct_pose}.

For each pose condition, the skull CT volume was transformed within the fixed array coordinate system. Orthogonal XZ and YZ CT slices intersecting the imaging planes were then extracted and passed through the full ultrasound image-formation workflow. The RF simulation, beamforming, B-mode formation, image enhancement, skull-edge extraction, and 3D feature reconstruction were rerun independently for each pose. Thus, each registration method was evaluated using image-derived features generated under the same pose-specific acoustic geometry rather than reusing a single nominal ultrasound dataset.

\subsection{Ultrasound image simulation and B-mode reconstruction}

The US-Cross images were simulated from the nine 64-element CMUT imaging arrays using the Field II MATLAB toolbox \cite{jensen1992calculation,jensen1996field}. Each array operated at 3.5 MHz with 90\% fractional bandwidth. For each XZ or YZ imaging plane, RF data were generated using five plane-wave transmissions steered from -20$^{\circ}$ to +20$^{\circ}$, and receive beamforming was performed with an F-number of 1. A custom pixel-based plane-wave compounding beamformer reconstructed the B-mode images by calculating transmit and receive delays for each pixel, interpolating delayed RF samples, and combining the compounded data using delay-multiply-and-sum (DMAS) beamforming \cite{matrone2015delay,matrone2016plane}. DMAS was used to emphasize coherent skull-interface returns relative to incoherent background signals. The compounded images were further processed using axial Wiener filtering to reduce shadow-like smearing, two-dimensional Gaussian smoothing to suppress speckle-scale fluctuations, and intensity adjustment to improve skull-edge contrast before feature extraction. Representative reconstructed B-mode images and processed skull-edge maps are shown in Fig.~\ref{fig:bmode_edges}.

\subsection{Skull-edge extraction and 3D US-Cross reconstruction}

Processed B-mode images were converted into geometric registration features using an automated edge-extraction pipeline. A bright-trace search first identified the dominant high-intensity response along each axial line, providing an initial estimate of the leading skull interface. This trace was refined using Canny-based edge detection to align points with local image-gradient maxima rather than the center of the bright band \cite{canny1986computational}. Outliers were removed, and the remaining points were regularized using spline smoothing to obtain a continuous skull contour. The refined XZ and YZ edges were then back-projected into the fixed array coordinate system using the known imaging-plane geometry, forming an orthogonal 3D US-Cross profile. For analysis, the extracted XZ and YZ features were combined into an orthogonal XZ+YZ US-Cross profile to test whether the two-plane configuration provided stable registration.

\subsection{Pulse-echo point generation from the therapeutic array}

\begin{figure}[!ht]
\centering
\includegraphics[width=0.47\textwidth]{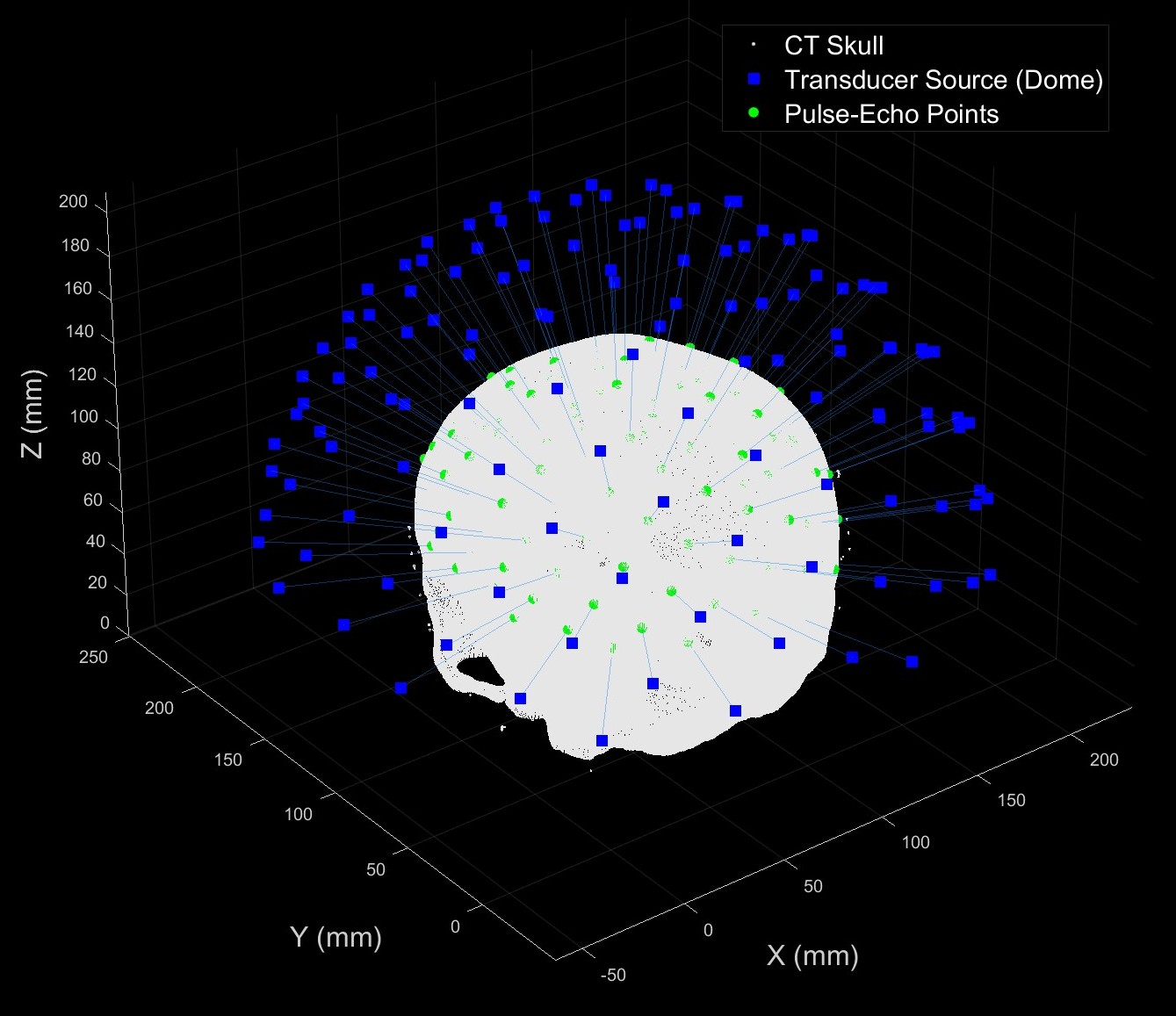}
\caption{Simulated therapeutic pulse-echo points generated by the 128-element hemispherical array and visualized using the ray-tracing method.}
\label{fig:pulse_echo}
\end{figure}

Sparse therapeutic pulse-echo observations were simulated using a first-hit ray model. For each of the randomly positioned 128 therapeutic elements, the element normal defined the direction toward the CT outer-surface model, and the first ray-surface intersection was selected as the simulated pulse-echo return (Fig.~\ref{fig:pulse_echo}). This produced a sparse global point cloud analogous to first-return skull localization measurements used in prior hemispherical ultrasound-to-CT registration systems \cite{oreilly2016registration,deng2022ultrasoundguided}. These points provide broad skull coverage and global pose constraint, but because the skull is smooth and approximately hemispherical, sparse pulse-echo data alone can remain ambiguous, particularly for rotation.

\subsection{Point-cloud preprocessing}

Before registration, all CT and ultrasound-derived point sets were expressed in the fixed array coordinate system. The CT outer surface was downsampled when needed to reduce computation while preserving skull curvature. Invalid therapeutic first-hit returns were excluded, and US-Cross points were resampled along the extracted contours so that the dense local imaging features did not dominate the objective solely because of higher point count. The same preprocessing protocol was applied across all pose conditions and registration modes.

\begin{figure*}[!ht]
\centering
\includegraphics[width=0.82\textwidth]{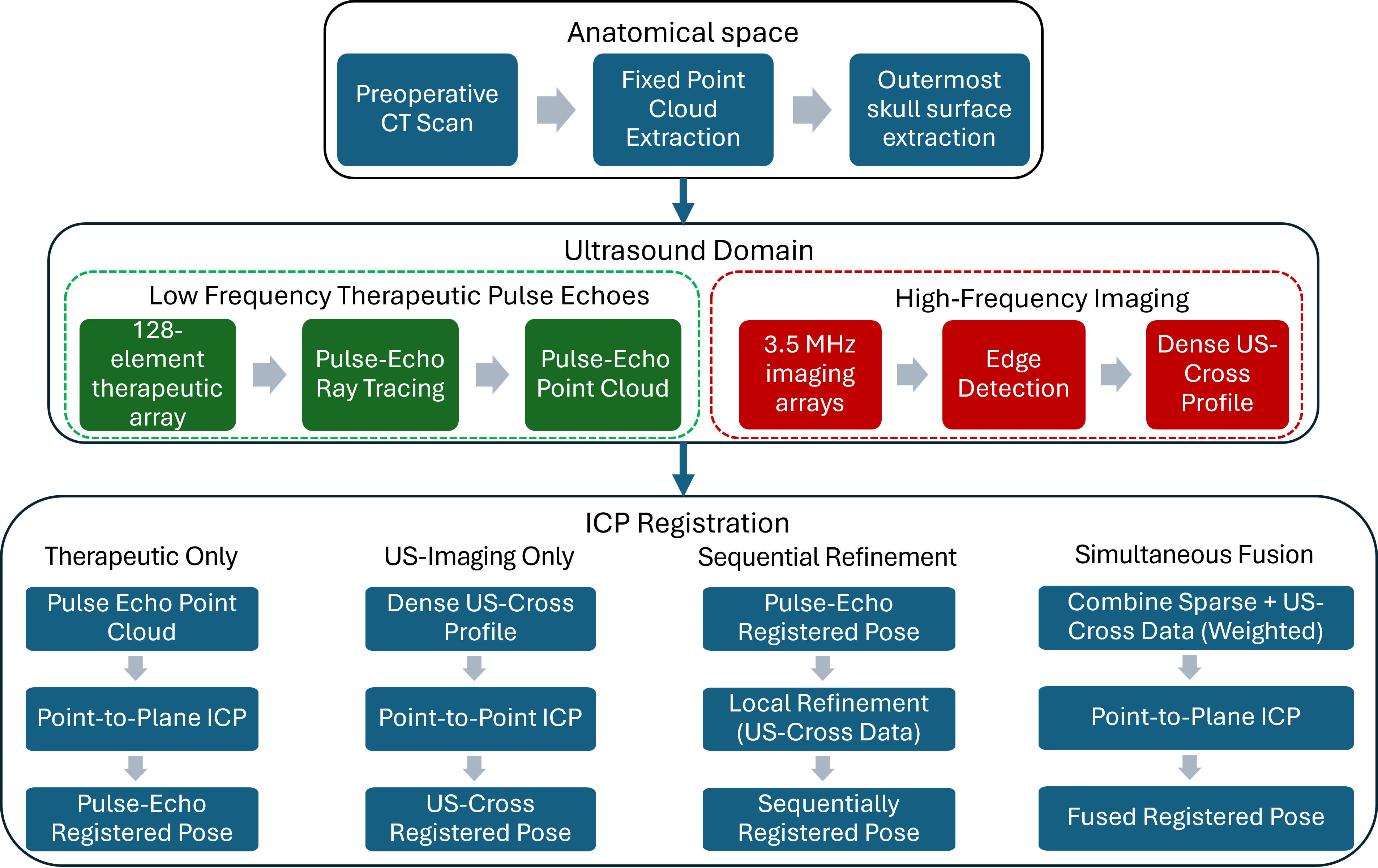}
\caption{Registration pipeline overview: Therapeutic only, US-imaging only, Sequential refinement, and Fused registration modes.}
\label{fig:pipeline}
\end{figure*}

\subsection{Rigid point-cloud registration strategy}

Registration was formulated as six-degree-of-freedom rigid alignment between ultrasound-derived source features and the CT-derived outer skull surface. Iterative closest point (ICP) was used as the main optimization framework because it is widely used for rigid 3D shape registration and has been adopted in prior ultrasound-to-CT skull registration studies \cite{oreilly2016registration,besl1992method,deng2022ultrasoundguided}. Point-to-plane ICP was used when sufficient surface support and reliable CT normals were available, because it is well suited to smooth skull surfaces \cite{chen1992object,rusinkiewicz2001efficient}.

For source point $p_i$, corresponding CT target point $q_i$, and CT surface normal $n_i$, the point-to-plane ICP objective was written as:

\begin{equation}
E(R,t)=\sum_i \left[\left(Rp_i+t-q_i\right)\cdot n_i\right]^2 .
\label{eq:point_to_plane}
\end{equation}

where R and t are the rotation matrix and translation vector to be estimated. Point-to-point ICP was retained for line-dominated US-imaging-only cases, where the extracted ultrasound features are sparse orthogonal curves rather than a full local surface and tangent-plane constraints can become under-constrained. Direct closest-point alignment therefore provides a more stable local refinement for these features.


\subsection{Registration modes}

Four registration modes were evaluated on identical pose-specific inputs, using the same CT target, coordinate frame, initialization, and convergence settings unless otherwise specified (Fig.~\ref{fig:pipeline}).

\noindent\textit{\textbf{Therapeutic only:}} The sparse first-hit point cloud from the 128-element therapeutic array was registered directly to the CT outer skull surface. This mode tested the accuracy achievable using the broad hemispherical therapeutic aperture alone. Point-to-plane ICP effectively utilizes the smooth surface normals of CT skull models to enhance convergence rates, provided the input data is clean. Conversely, when processing degraded, noisy, or sparse returns, point-to-point alignment offers a more stable alternative by bypassing the computational instabilities associated with local tangent-plane residuals \cite{rusinkiewicz2001efficient}.

\noindent\textit{\textbf{US-imaging only:}} The reconstructed high-frequency US-Cross features were registered to the CT surface without the pulse-echo point cloud. Because the US-Cross consists of orthogonal curves rather than a dense 3D surface, this mode was implemented with point-to-point logic to avoid surface-sliding ambiguity and under-constrained point-to-plane behavior along line-like features.

\noindent\textit{\textbf{Sequential refinement:}} The sequential mode was implemented as a hierarchical global-to-local hand-off. It begins with an initial transform $T_{\mathrm{init}}$ in $\mathrm{SE}(3)$, representing the manual or GUI-assisted starting pose available before automatic optimization. This transform is defined by the initial rotation $R_{\mathrm{init}}$ and translation $t_{\mathrm{init}}$:

\begin{equation}
T_{\mathrm{init}} =
\begin{bmatrix}
R_{\mathrm{init}} & t_{\mathrm{init}} \\
0^{\top} & 1
\end{bmatrix}, \qquad T_{\mathrm{init}} \in \mathrm{SE}(3).
\label{eq:tinit}
\end{equation}

Starting from $T_{\mathrm{init}}$, the therapeutic point cloud was first registered to the CT surface to obtain a global therapeutic transform $T_{\mathrm{PE}}$. This estimate then initialized a second registration stage using the US-Cross features, producing a local refinement transform $T_{\mathrm{USCross}}$. The final sequential transform was

\begin{equation}
T_{\mathrm{sequential}} = T_{\mathrm{USCross}} T_{\mathrm{PE}} T_{\mathrm{init}} .
\label{eq:tseq}
\end{equation}

This strategy is designed to prevent the higher frequency US-Cross data from being asked to solve a large global misalignment. The pulse-echo stage provides the global lock, while the US-Cross stage supplies the localized geometric correction needed for higher precision.

\noindent\textit{\textbf{Fused registration:}} The fused mode combined pulse-echo and US-Cross features in a single weighted optimization. This approach estimates one rigid transform by minimizing a joint objective,

\begin{equation}
\begin{split}
E_{\mathrm{fused}}(R,t) = &\sum_{i \in \mathrm{PE}} w_D
\left[\left(Rp_i+t-q_i\right)\cdot n_i\right]^2 \\
&+ \sum_{j \in \mathrm{USCross}} w_L
\left[\left(Rp_j+t-q_j\right)\cdot n_j\right]^2 .
\end{split}
\label{eq:fused_cost}
\end{equation}

where PE denotes the therapeutic point set, USCross denotes the US-Cross point set, $p_i$ and $p_j$ are ultrasound-derived source points, $q_i$ and $q_j$ are nearest-neighbor CT target points, $n_i$ and $n_j$ are the corresponding CT surface normals, and $w_D$ and $w_L$ control the relative influence of the therapeutic array and high-frequency imaging terms. Point-count normalization and downsampling were used so that the dense US-Cross curves did not dominate the cost function solely because they contained more samples than the sparse therapeutic point cloud. The final fused transform was:

\begin{equation}
T_{\mathrm{fused}} = T_{\mathrm{USPE}} T_{\mathrm{init}} .
\label{eq:tfused}
\end{equation}

where $T_{\mathrm{USPE}}$ is the rigid transformation estimated from the combined therapeutic and US-Cross feature set. In contrast to the sequential mode, the fused mode allows the optimizer to balance global hemispherical constraints and local high-frequency features in the same iteration. This makes it the more conservative option when isolated modality estimates are unstable, while the sequential mode remains attractive when the therapeutic stage already provides a reliable global initialization.

\subsection{Initialization, convergence, and fair-comparison controls}

Because ICP is a local optimizer, all registration modes were initialized with the same $T_{\mathrm{init}}$ for each pose condition. The CT target, pose definition, imaging geometry, image-processing parameters, downsampling rules, and convergence criteria were also held constant across methods unless explicitly defined as part of a specific registration mode. This ensured that differences in displacement or rotation error reflected the available spatial constraints and optimization strategy rather than different initial conditions or pre-processing choices.

\begin{figure*}[!ht]
\centering
\includegraphics[width=0.82\textwidth]{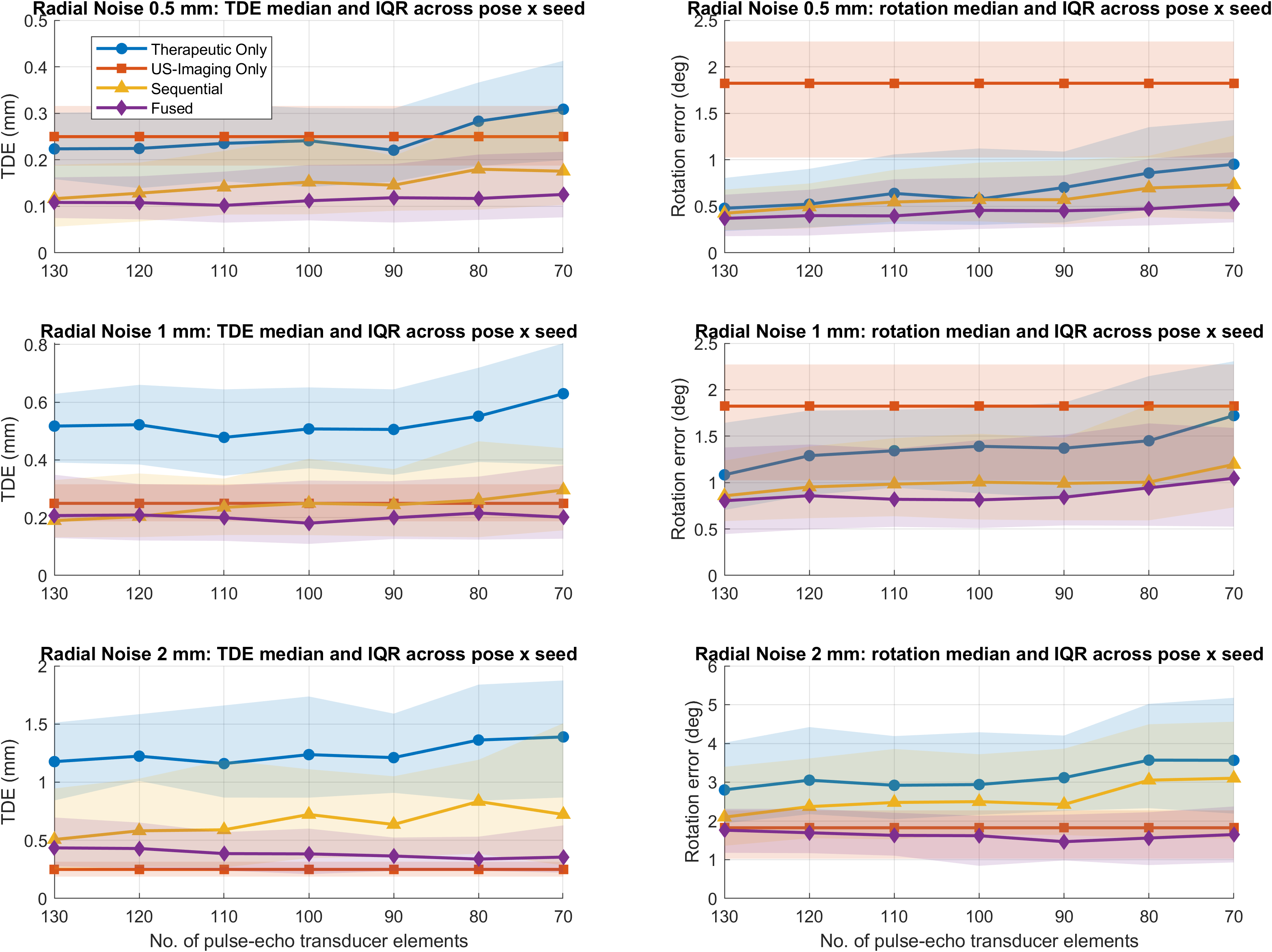}
\caption{Main performance summary: TDE and rotation error across the 25 pose conditions and five radial-noise seeds for all four registration modes.}
\label{fig:main_performance}
\end{figure*}

\subsection{Error metrics and statistical analysis}

Registration accuracy was quantified by comparing the estimated transform with the known ground-truth skull-array transform used to generate each simulated pose. For robustness tests, the ground-truth pulse-echo profile was defined by the first-hit points before stochastic radial noise was added. Five randomized radial-noise seeds were used for each of the 25 controlled poses, yielding 125 pose-seed realizations for each radial-noise and active-element setting.

The residual transform between the estimated and ground-truth poses was expressed as an error rotation $R_{\mathrm{err}}$ and error translation $t_{\mathrm{err}}$. Translational accuracy was reported as total displacement error (TDE), calculated as the Euclidean norm of the residual translation:

\begin{equation}
\mathrm{TDE}=\|t_{\mathrm{err}}\|_2=\sqrt{\Delta x^2+\Delta y^2+\Delta z^2} .
\label{eq:tde}
\end{equation}

Angular accuracy was reported as total rotation error, calculated from the trace of the residual rotation matrix:

\begin{equation}
\theta_{\mathrm{rot}}=\mathrm{rad2deg}\left[\arccos\left(\frac{\mathrm{Tr}(R_{\mathrm{err}})-1}{2}\right)\right] .
\label{eq:rot_error}
\end{equation}

Performance was summarized using median, interquartile range, mean, standard deviation, and threshold-based success rates for TDE $\leq$ 0.5, 1, and 2 mm and rotation error $\leq$ 0.5$^{\circ}$, 1$^{\circ}$, and 2$^{\circ}$. Because all registration modes were evaluated on matched pose-seed conditions, statistical comparisons were paired. Friedman tests were used to assess overall method differences, followed by pairwise Wilcoxon signed-rank tests with Holm correction. The pose-seed condition was the unit of analysis; statistical significance therefore reflects method behavior for the tested anatomy and pose space rather than inter-subject generalization.

\subsection{Secondary robustness benchmark}

A secondary robustness analysis evaluated registration stability under degraded therapeutic pulse-echo sampling. Radial range perturbations were added to model uncertainty in first-return localization, and the number of active therapeutic elements was reduced to simulate partial element dropout or poor acoustic reflection due to large incidence angles. The five radial-noise seeds avoided drawing conclusions from a single spatial noise realization. This analysis tested whether US-Cross features provided an additional geometric anchor when the sparse global pulse-echo constraint became noisier or less complete.

\section{Results}

\begin{table*}[t]
\caption{Overall translational and rotational registration performance across all matched pose--seed realizations, radial-noise levels, and active therapeutic-element conditions.}
\label{tab:summary}
\centering
\scriptsize
\resizebox{\textwidth}{!}{%
\begin{tabular}{lcccccccccc}
\hline
\multirow{2}{*}{Method} &
\multicolumn{5}{c}{Total displacement error (TDE)} &
\multicolumn{5}{c}{Rotation error} \\
\cline{2-6} \cline{7-11}
&
\begin{tabular}{c}Median [IQR]\\(mm)\end{tabular} &
\begin{tabular}{c}$\leq$0.5 mm\\(\%)\end{tabular} &
\begin{tabular}{c}$\leq$1 mm\\(\%)\end{tabular} &
\begin{tabular}{c}$\leq$2 mm\\(\%)\end{tabular} &
\begin{tabular}{c}Mean\\(mm)\end{tabular} &
\begin{tabular}{c}Median [IQR]\\(deg)\end{tabular} &
\begin{tabular}{c}$\leq$0.5 deg\\(\%)\end{tabular} &
\begin{tabular}{c}$\leq$1 deg\\(\%)\end{tabular} &
\begin{tabular}{c}$\leq$2 deg\\(\%)\end{tabular} &
\begin{tabular}{c}Mean\\(deg)\end{tabular} \\
\hline
Therapeutic only &
0.518 [0.278--0.961] &
48.3 &
76.3 &
95.1 &
0.715 &
1.440 [0.749--2.505] &
16.5 &
35.2 &
65.1 &
1.929 \\

US-imaging only &
0.250 [0.188--0.316] &
84.0 &
100.0 &
100.0 &
0.286 &
1.823 [1.024--2.272] &
4.0 &
16.0 &
60.0 &
1.688 \\

Sequential &
0.299 [0.147--0.526] &
73.8 &
89.0 &
97.3 &
0.443 &
1.034 [0.570--2.093] &
21.3 &
50.0 &
73.7 &
1.609 \\

Fused &
0.271 [0.169--0.420] &
85.1 &
98.4 &
100.0 &
0.278 &
0.904 [0.432--1.639] &
29.9 &
53.8 &
84.3 &
1.130 \\
\hline
\end{tabular}%
}
\end{table*}

\subsection{Registration performance across the controlled pose matrix}

Fig.~\ref{fig:main_performance} summarizes registration performance across the controlled pose matrix. For each radial-noise and active-element setting, median and interquartile regions were calculated across 25 poses and five randomized radial-noise seeds, yielding 125 matched pose-seed realizations. The registration mode consistently affected both translational and rotational accuracy, with Friedman omnibus tests showing significant method-dependent differences across all tested noise and active-element conditions.

Therapeutic only produced the least stable translational performance, with an overall median TDE of 0.518 mm [0.278--0.961] and mean TDE of 0.715 mm (Table~\ref{tab:summary}). Incorporating US-Cross imaging information reduced both the central tendency and spread of TDE. US-imaging only achieved the lowest median TDE, 0.250 mm [0.188--0.316], followed by Fused registration, 0.271 mm [0.169--0.420], and Sequential refinement, 0.299 mm [0.147--0.526]. These results demonstrate that local US-Cross contours effectively lock the position of the visible skull interface; however, they do not establish ultrasound as an independent modality for solving the complete spatial pose.

Rotational accuracy showed a different ranking (Table~\ref{tab:summary}). Fused registration achieved the lowest median rotation error, 0.904$^{\circ}$ [0.432--1.639], followed by Sequential refinement, 1.034$^{\circ}$ [0.570--2.093], Therapeutic only, 1.440$^{\circ}$ [0.749--2.505], and US-imaging only, 1.823$^{\circ}$ [1.024--2.272]. Thus, while local high-frequency contours were highly effective for translational correction, the broad hemispherical pulse-echo aperture remained important for constraining global orientation. Overall, Fused registration provided the most balanced performance across displacement and rotation metrics.

\begin{figure*}[!ht]
\centering
\subfloat[]{\includegraphics[width=0.47\textwidth]{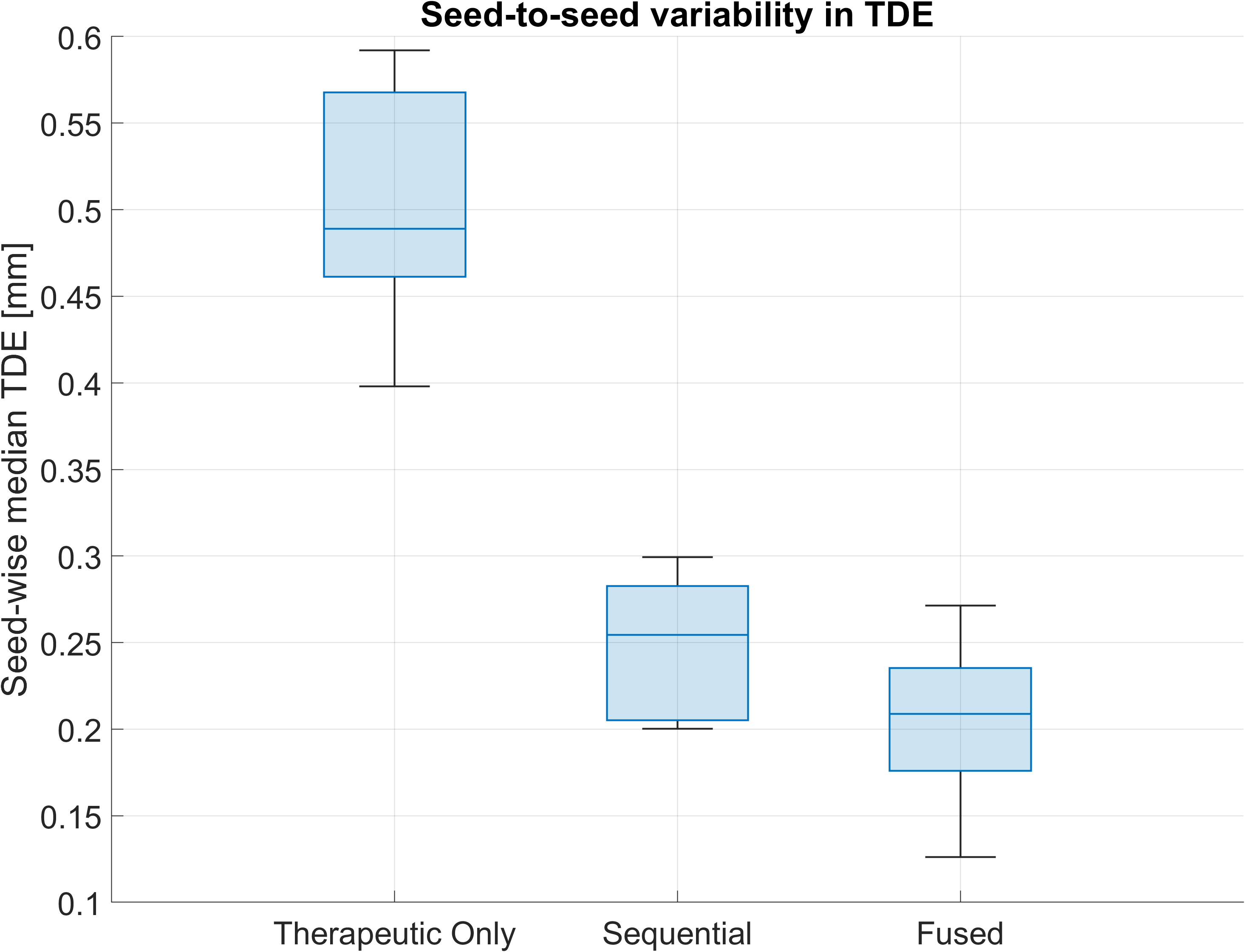}}\hfill
\subfloat[]{\includegraphics[width=0.47\textwidth]{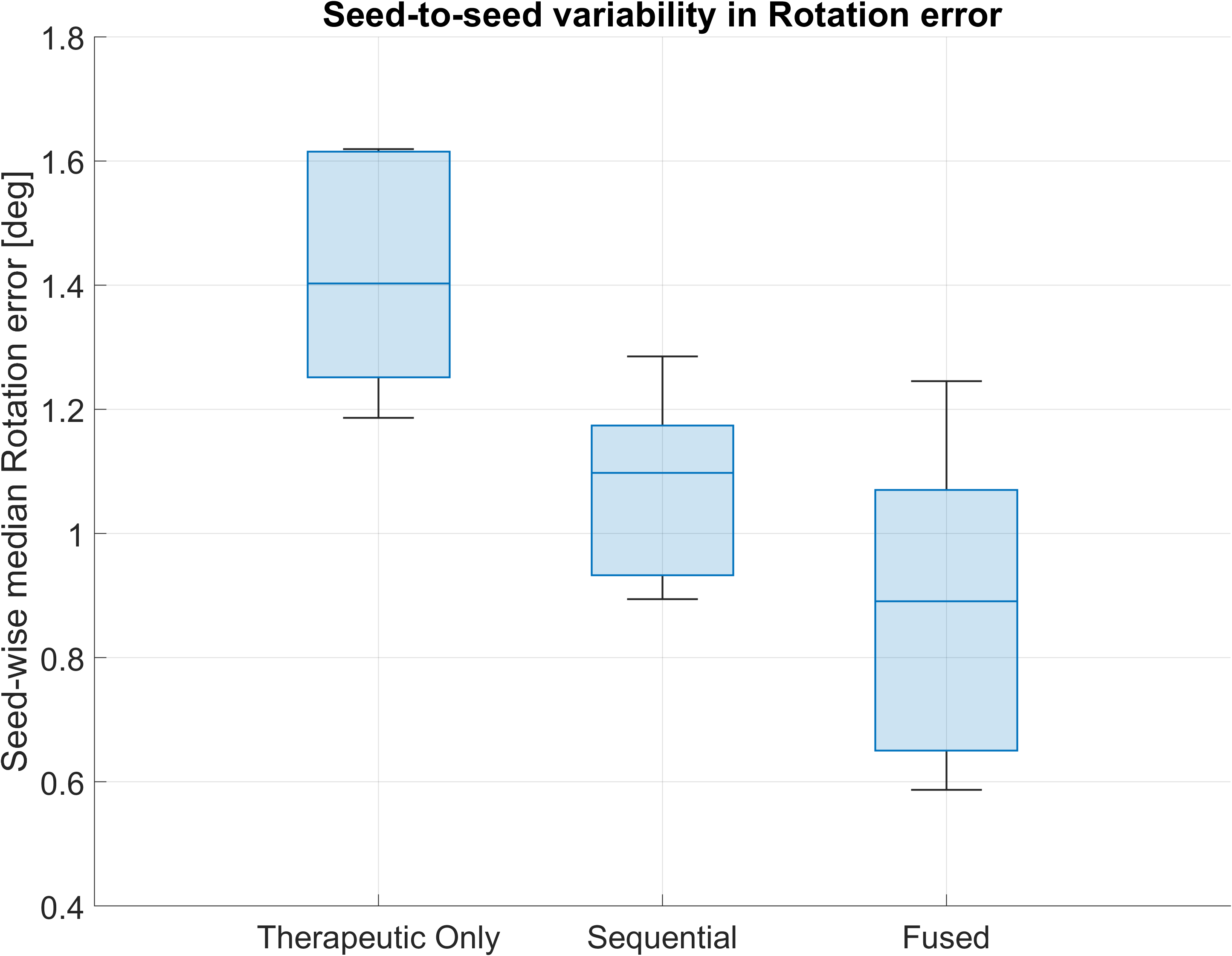}}
\caption{Seed-to-seed variability in TDE and rotation error.}
\label{fig:seed_variability}
\end{figure*}

Fig.~\ref{fig:seed_variability} further shows how these trends varied across the five randomized radial-noise seeds. Therapeutic only exhibited the largest seed-to-seed variability, particularly as radial pulse-echo uncertainty increased, indicating that sparse therapeutic first-return points are sensitive to the spatial distribution of measurement or element-placement error. Sequential refinement reduced this variability when the pulse-echo stage provided a reasonable global initialization, but its performance still partly depended on the stability of the first-stage estimate. Fused registration showed more consistent TDE and rotation-error behavior across seeds because the global pulse-echo points and local US-Cross contours were optimized together. US-imaging only does not show seed-to-seed variability in this figure because the randomized radial noise was applied only to the therapeutic pulse-echo measurements; the US-Cross edge profiles were deliberately held fixed to represent a best-case local imaging constraint. Therefore, the absence of US-Cross IQR should not be interpreted as experimental noise-free imaging, but rather as a controlled reference for evaluating how stable imaging-derived edges influence the other registration modes.

\subsection{Success-rate analysis}

Threshold-based success rates provided a clinically interpretable measure of registration robustness (Table ~ \ref{tab:summary}, and Fig.~\ref{supp-fig:S1-success_rates}). For translational accuracy, Therapeutic only achieved TDE $\leq$0.5 mm in 48.3\% of conditions, $\leq$1 mm in 76.3\%, and $\leq$2 mm in 95.1\%. In comparison, Fused registration reached 85.1\%, 98.4\%, and 100.0\% at the same thresholds, while Sequential refinement reached 73.8\%, 89.0\%, and 97.3\%. US-imaging only achieved 84.0\%, 100.0\%, and 100.0\%, respectively, reflecting the strong local translational constraint provided by the high-frequency skull-edge profiles.

Rotation thresholds showed a different pattern. Fused registration produced the highest rotational success rates, with 29.9\% of conditions below 0.5$^{\circ}$, 53.8\% below 1$^{\circ}$, and 84.3\% below 2$^{\circ}$. Sequential refinement reached 21.3\%, 50.0\%, and 73.7\%, while Therapeutic only reached 16.5\%, 35.2\%, and 65.1\%. US-imaging only had the weakest rotational threshold performance, reaching 4.0\%, 16.0\%, and 60.0\%. These results reinforce the main finding that local US-Cross features strongly improve displacement accuracy, but full rotational stability requires global pulse-echo anchoring, with Fused registration providing the most balanced threshold performance.

\begin{figure*}[!t]
\centering
\subfloat[]{\includegraphics[width=0.47\textwidth]{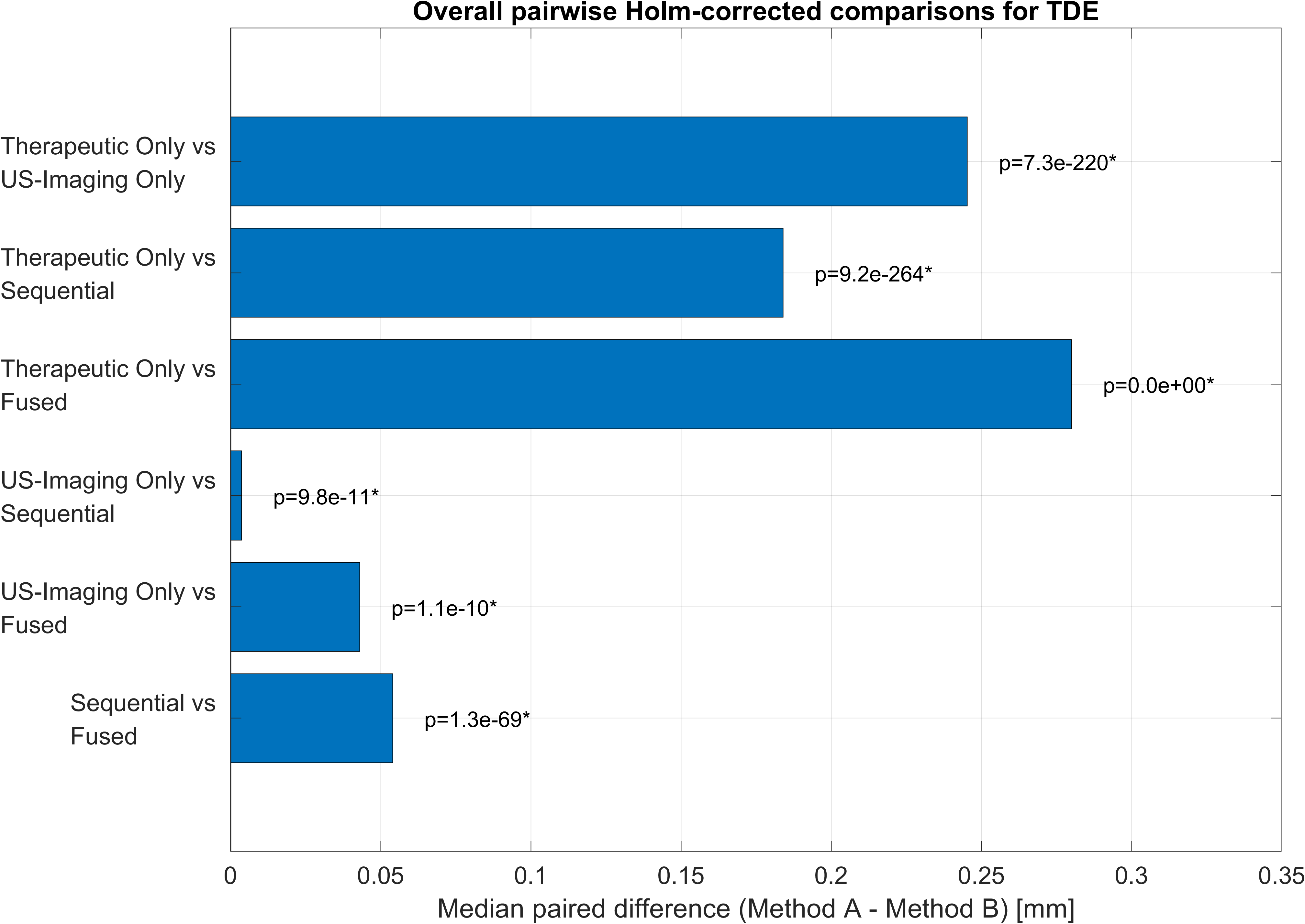}}\hfill
\subfloat[]{\includegraphics[width=0.47\textwidth]{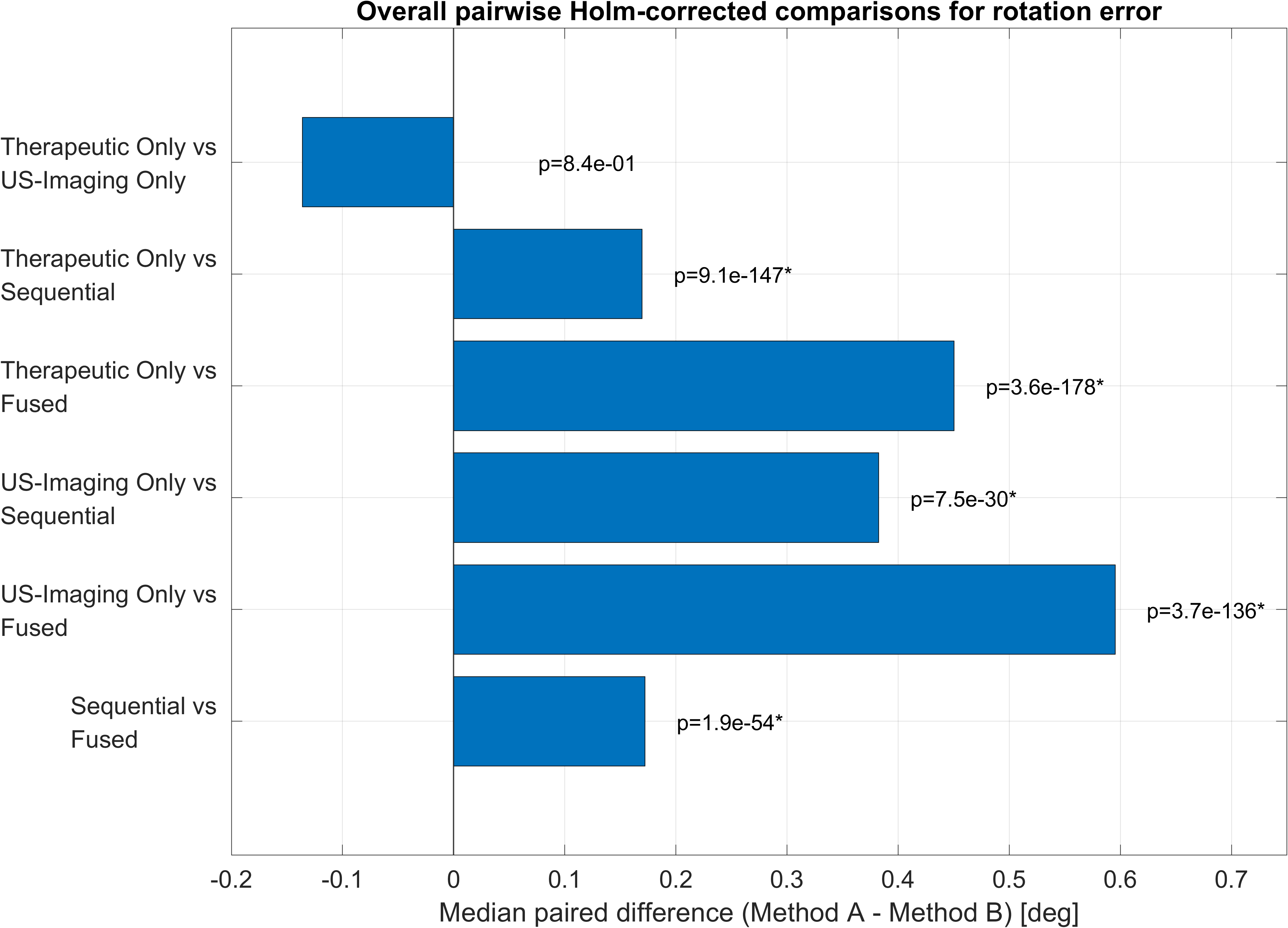}}
\caption{Holm-corrected pairwise Wilcoxon comparisons for TDE and rotation error. Positive values indicate that Method B had lower median error than Method A for the labeled comparison.}
\label{fig:pairwise}
\end{figure*}

Pairwise Wilcoxon signed-rank tests with Holm correction were used to identify which registration modes differed under matched pose-seed conditions (Fig.~\ref{fig:pairwise}). For TDE, all pairwise comparisons remained significant after correction. Relative to Therapeutic only, the median TDE was reduced by 0.268 mm for US-imaging only, 0.219 mm for Sequential refinement, and 0.247 mm for Fused registration, confirming that inclusion of high-frequency US-Cross information significantly improved translational accuracy.

For rotation error, Fused and Sequential registration both significantly improved upon Therapeutic only, whereas Therapeutic only and US-imaging only were not significantly different after Holm correction. Fused registration also significantly reduced rotation error relative to Sequential refinement and US-imaging only.

\subsection{Omnibus and pose-dependent behavior}

The Friedman omnibus evidence maps further show that registration performance was method-dependent across the full range of radial pulse-echo noise and active therapeutic-element counts (Supplementary Fig.~\ref{supp-fig:s2_friedman}). For TDE, all tested conditions showed significant differences among the four registration modes, with the strongest evidence observed at the highest radial-noise level, where p-values reached the order of $10^{-43}$--$10^{-49}$. Rotation error also exhibited significant method-dependent differences across all tested conditions.

Pose-wise analysis confirmed that the benefit of fusion was not limited to a single configuration (Supplementary Fig.~\ref{supp-fig:s3_posewise}). Across the 25 center-position and angular-condition combinations, Fused registration reduced median TDE relative to Therapeutic only in all tested poses. The largest improvement occurred for C2A0, where the median TDE reduction reached 0.47 mm. 

\section{Discussion}

This study demonstrates that integrated US-Cross imaging information improves CT-to-ultrasound skull registration relative to localization using sparse therapeutic array pulse-echo data alone. The benefit was clearest for translational accuracy: Therapeutic only produced a median TDE of 0.518 mm, whereas the US-imaging-informed modes reduced the median TDE to 0.250--0.299 mm and increased the fraction of cases below submillimeter thresholds. These results support the central premise that localized high-frequency CMUT imaging can provide useful geometric constraints for MRI-free skull registration in a hemispherical tFUS system.

The results also clarify the distinct roles of local and global acoustic information. US-imaging only achieved the lowest median TDE and 100\% success below the 1-mm and 2-mm translational thresholds, but it had the worst rotational threshold performance. This is physically consistent with the geometry of the problem: orthogonal high-frequency skull contours provide strong local information about the position of the outer skull surface, but line-like local features alone do not fully constrain the global orientation of a smooth, approximately hemispherical skull. Therefore, the US-imaging-only mode should not be viewed as a replacement for global therapeutic localization. Its main value is as a high-resolution local anchor that complements the information obtained from broad-aperture therapeutic measurements.


The seed-to-seed analysis further supports this interpretation. Randomized radial-noise seeds were applied only to the therapeutic pulse-echo measurements, while the US-Cross edge profiles were deliberately held fixed to represent a best-case local imaging constraint. Consequently, US-imaging only did not show seed-to-seed IQR in the variability analysis. This should not be interpreted as evidence that US-imaging would be noise-free experimentally. In a physical implementation, US-imaging performance would also depend on image SNR, edge-detection uncertainty, acoustic coupling, array calibration, and local skull incidence angle. Instead, the controlled comparison shows that when local imaging-derived edges are reliable, they can stabilize registration against variability in sparse therapeutic pulse-echo measurements.

Sequential and Fused registration represent two complementary strategies for combining these constraints. Sequential refinement is intuitive: pulse-echo registration first moves the skull estimate into the correct global neighborhood, and US-Cross features then refine the local alignment. This can be effective when the first-stage pulse-echo estimate is reliable, but it inherits the convergence basin of that initial stage. Fused registration reduces this dependence by optimizing global pulse-echo points and local US-Cross contours simultaneously. In the multi-seed analysis, Fused registration provided the best overall rotational performance and the strongest combined threshold behavior, reaching 98.4\% success below 1-mm TDE, 100.0\% below 2-mm TDE, and 84.3\% below 2$^{\circ}$ rotation error. This makes Fused registration the most conservative option when the initial geometry or pulse-echo data quality is uncertain.

Our findings demonstrate that while local US-Cross imaging features strongly constrain the position of the visible skull interface, ultrasound imaging alone does not yield the optimal full-pose solution. Rather, Fused registration provides the most robust full-pose estimation by balancing these localized imaging details with the overall therapeutic geometry. This dynamic is evident in the structured pattern of rotation errors, which reflects the distinct ways that sparse global pulse-echo points and local US-Cross contours constrain spatial orientation. Notably, our data indicate that the choice of registration strategy becomes increasingly critical as therapeutic pulse-echo localization uncertainty rises. Because simultaneous sparse-plus-dense registration improved translational accuracy broadly across the entire controlled pose matrix, these results strongly support the use of integrated high-frequency imaging as a stabilizing geometric constraint rather than merely a pose-specific correction.

The significance of a given registration error depends strongly on the therapy frequency and corresponding focal dimensions. O'Reilly et al. used an 11-MHz array for skull localization but evaluated the downstream impact of the registration error for a 230-kHz therapeutic system, where 0.9 $\pm$ 0.2~mm displacement and 0.8 $\pm$ 0.4$^{\circ}$ rotation errors were predicted to produce approximately 1.0 $\pm$ 0.2~mm targeting error \cite{oreilly2016registration}. In their simulations, an error of this scale represented approximately 26\% of the lateral $-6$-dB focal width at 230~kHz but approximately 74\% at 650~kHz, emphasizing that the same geometric error becomes more consequential as the focal volume contracts \cite{oreilly2016registration}. Deng et al. likewise used 11-MHz skull-localization elements, while the therapeutic array operated at 306 and 612~kHz; targeting errors remained approximately 0.9~mm at both therapy frequencies \cite{deng2022ultrasoundguided}. At the higher end, Chauvet et al. demonstrated millimetric end-to-end targeting accuracy with a 1-MHz MR-guided HIFU system in fresh cadaver heads \cite{chauvet2013targeting}. These results should not be interpreted as a single universal tolerance: the required precision also depends on target size, biological endpoint, steering distance, and the aberration-correction method. Nevertheless, the frequency dependence of focal dimensions increases the potential value of submillimeter registration for higher-frequency transcranial applications, although therapeutic beam simulations are required to establish the acceptable tolerance for any specific system and target.

Cross-modality comparisons also require care because different studies include different portions of the targeting error chain. The 0.9-mm registration error reported by O'Reilly et al. was obtained with ex vivo skullcaps without overlying scalp under controlled degassed-water coupling \cite{oreilly2016registration}, whereas optically tracked FUS systems have reported end-to-end TREs of approximately 2.4--3.3~mm when calibration, image registration, and targeting are included \cite{chaplin2019accuracy}. High-resolution MRI-to-CT registration in cranial stereotactic workflows can approach the 1-mm level, but values around 2~mm are also reported more generally \cite{putz2024quality}. In the present study, Fused registration achieved a median TDE of 0.271~mm and a median rotation error of 0.904$^{\circ}$. A simple lever-arm estimate indicates that a 0.904$^{\circ}$ residual rotation would produce approximately 0.47, 0.79, and 0.95~mm displacement at 30, 50, and 60~mm target depths, respectively. Combined in quadrature with the median Fused TDE and a residual skull-induced focal shift of approximately 0.3~mm reported after two-step correction in recent histotripsy/navigation studies \cite{lu2022twostep,choi2024navith}, the estimated median end-to-end shift would be on the order of 0.6--1.0~mm for 30--60~mm target depths. This estimate is encouraging but remains illustrative because true beam shift depends on skull anatomy, target location, aperture geometry, frequency, calibration error, and the specific aberration-correction method.


The main limitation of this study is that it remains a controlled technical evaluation on one skull anatomy; therefore, inter-subject anatomical variability was not assessed. The pose sweep and randomized pulse-echo seeds test skull-array pose sensitivity and method robustness, but further studies are needed to establish generalization across subjects. The US-Cross features were also simulated and were not perturbed by independent imaging noise, so the US-imaging-only result represents a best-case local edge constraint. In addition, the rigid transform between the therapeutic array and the CMUT imaging subarrays was assumed to be known exactly. A physical implementation will require calibration of this inter-array transform and experimental quantification of its translational and rotational uncertainty. As a scale reference rather than an estimate for the present hardware, O'Reilly et al. reported approximately 0.5 $\pm$ 0.1~mm displacement and 0.5$^{\circ}$ $\pm$ 0.2$^{\circ}$ rotation error associated with repeatable skull-frame placement in their experimental registration setup \cite{oreilly2016registration}.

The present simulations also omit overlying skin and scalp. These tissues may affect high-frequency image quality and pulse-echo localization through attenuation, scattering, coupling variability, and sound-speed differences, and may therefore represent an important confounding factor in vivo. O'Reilly et al. likewise identified the absence of scalp as a limitation of skull-surface pulse-echo registration and noted that full-aperture coupling through scalp required further study \cite{oreilly2016registration}. Future validation should progress from additional CT anatomies and physical phantoms to intact cadaveric heads or skull specimens with preserved soft tissue, followed by non-human-primate studies where appropriate. Such experiments should quantify skin/scalp effects, incidence-angle sensitivity, inter-array calibration error, and image-derived edge uncertainty. Finally, fully coupled patient-specific acoustic simulations should propagate the measured registration-error distributions into focal shift, focal pressure loss, and passive-acoustic-map localization error across relevant therapeutic frequencies. This end-to-end validation is needed to determine whether the proposed MRI-free registration framework preserves the therapeutic focus within the application-specific safety and targeting margin.

\bibliographystyle{IEEEtran}
\bibliography{references}

\end{document}


\maketitle

\section{Supplementary Results}

\begin{figure*}[!t]
\centering
\subfloat[]{\includegraphics[width=0.47\textwidth]{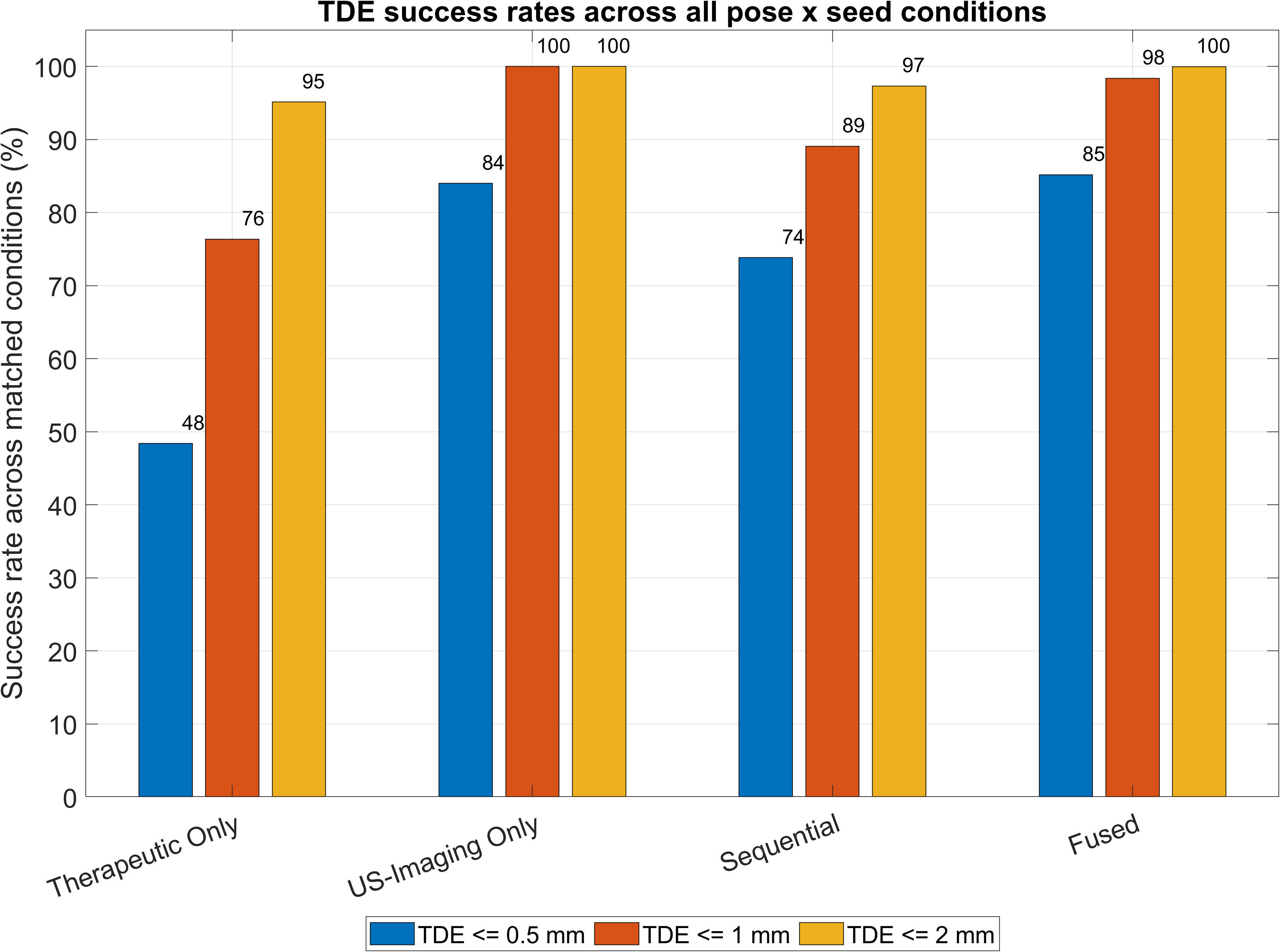}}\hfill
\subfloat[]{\includegraphics[width=0.47\textwidth]{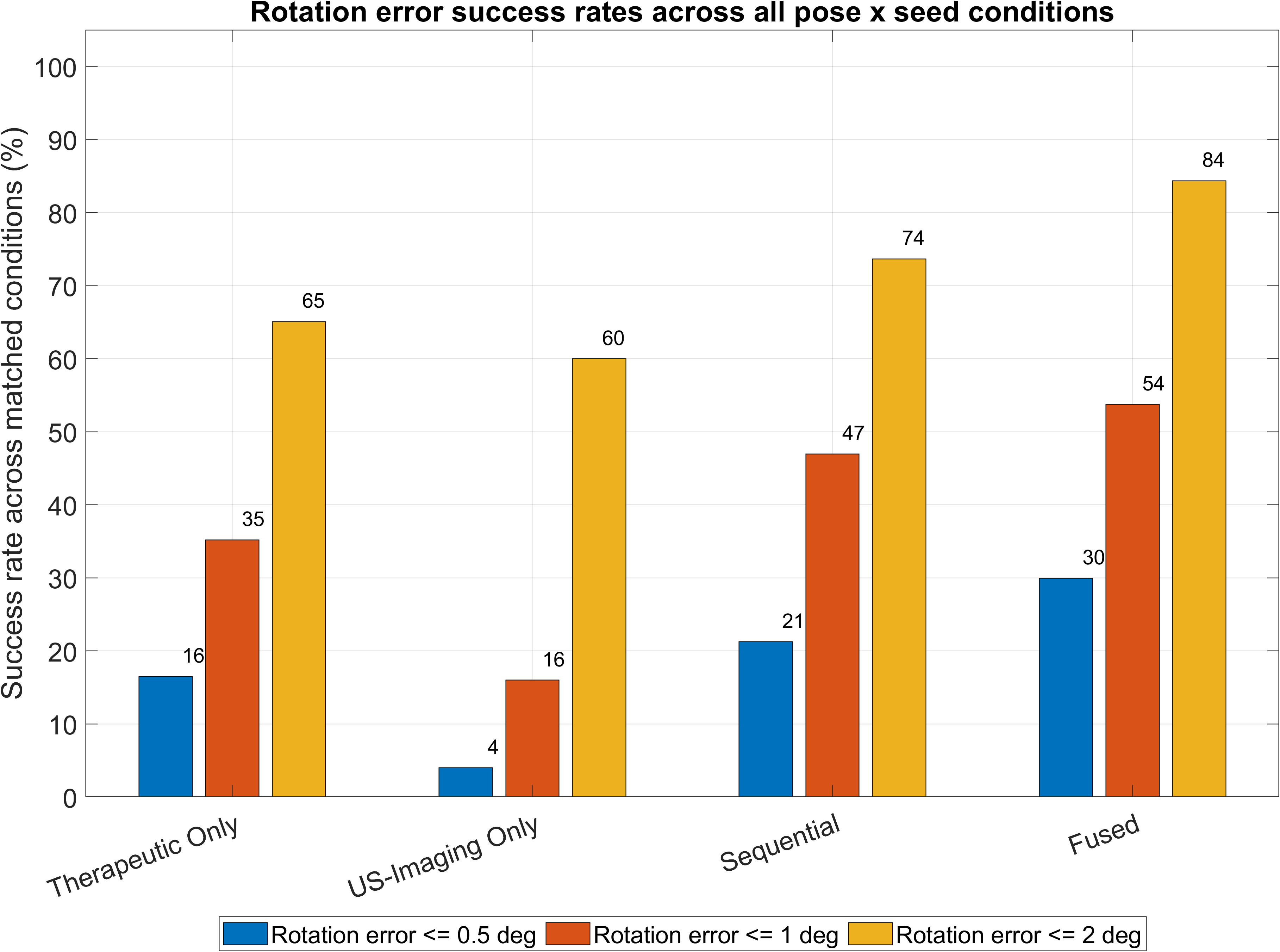}}
\caption{Translation-error and rotational-error success rates across all matched pose and seed conditions, radial-noise levels, and active-element counts using 0.5-mm, 1-mm, and 2-mm TDE thresholds in (a) and 0.5-degree, 1-degree, and 2-degree rotation thresholds in (b).}
\label{fig:S1-success_rates}
\end{figure*}

\begin{figure*}[!t]
\centering
\subfloat[TDE Friedman omnibus evidence map.]{
    \includegraphics[width=0.48\textwidth]{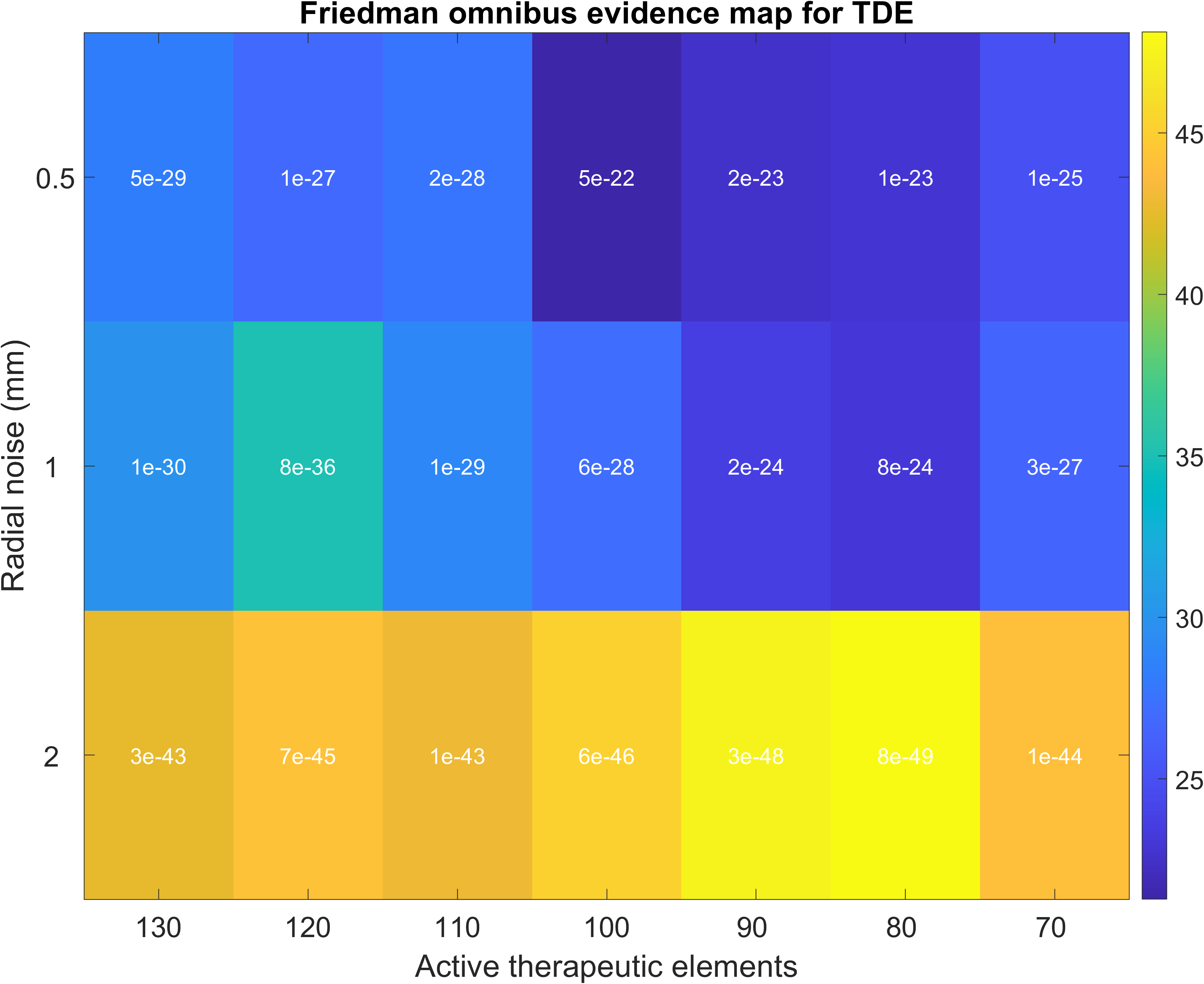}
    \label{fig:s2a}
}
\hfill
\subfloat[Rotation-error Friedman omnibus evidence map.]{
    \includegraphics[width=0.48\textwidth]{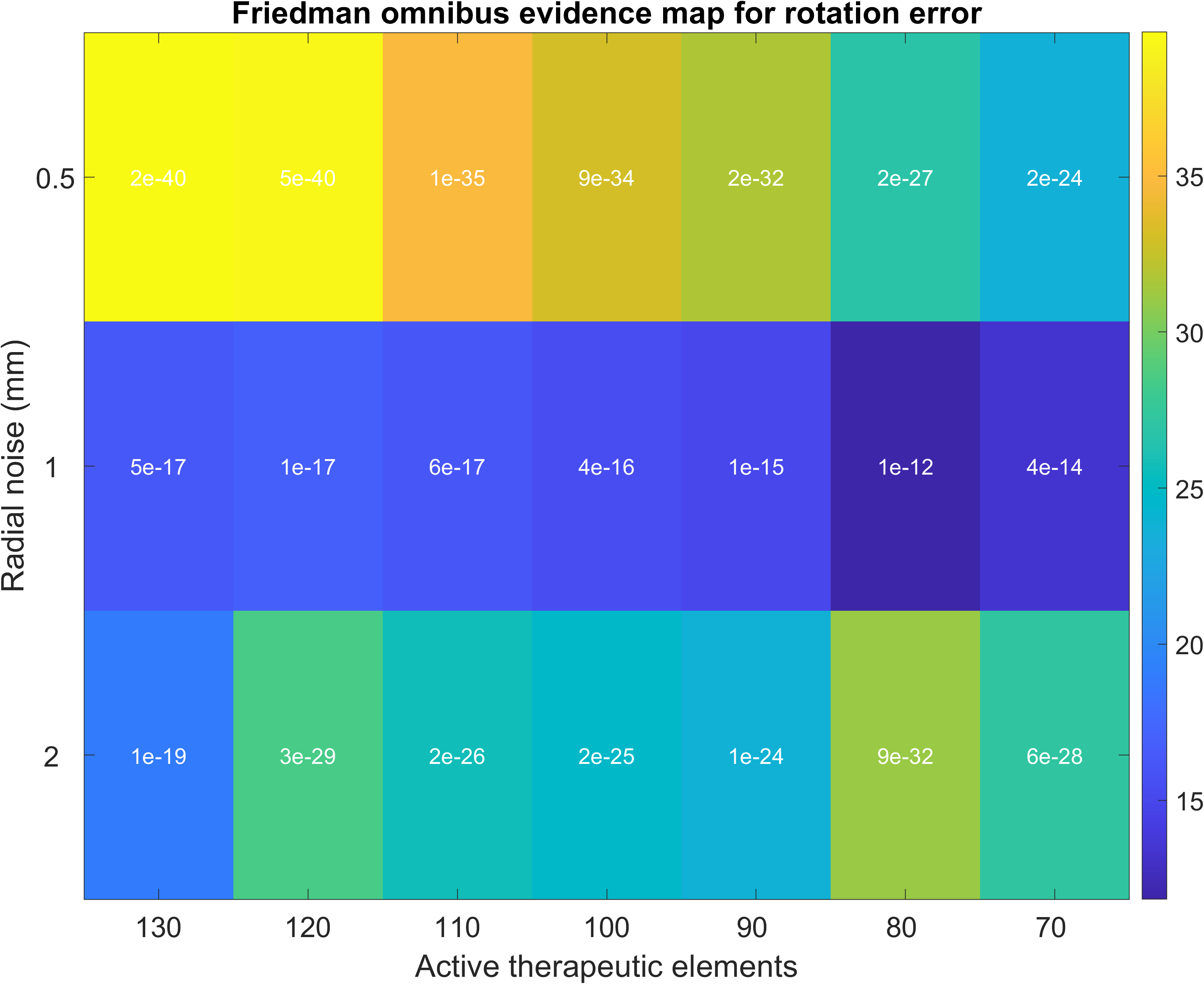}
    \label{fig:s2b}
}
\caption{Friedman omnibus evidence maps for TDE and rotation error across all poses and radial-noise seeds. These maps show that registration performance was method-dependent across the tested radial-noise and active therapeutic-element conditions.}
\label{fig:s2_friedman}
\end{figure*}

\begin{figure}[!t]
\centering
\includegraphics[width=\columnwidth]{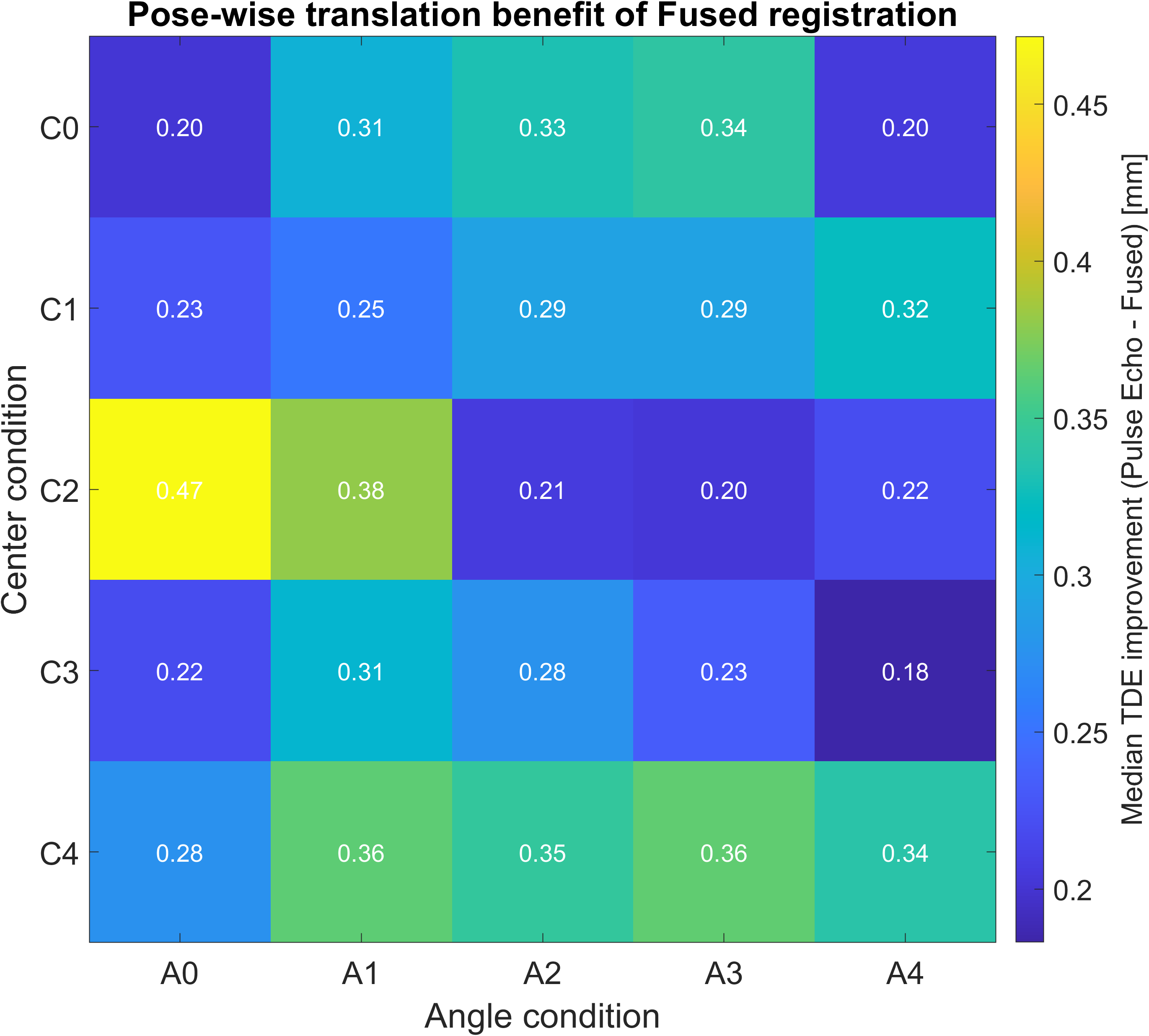}
\caption{Pose-wise median TDE improvement of Fused registration relative to Therapeutic only. Each cell corresponds to one center-position and angular-condition pair.}
\label{fig:s3_posewise}
\end{figure}

This supplementary material provides additional statistical evidence supporting the method-dependent behavior of the proposed CT-to-ultrasound skull registration framework. Supplementary Fig.~\ref{fig:S1-success_rates} is Threshold-based success rates and visualizes the summary preseneted at Table 1. Supplementary Fig.~\ref{fig:s2_friedman} shows the Friedman omnibus evidence maps across radial-noise and active-element conditions, and Supplementary Fig.~\ref{fig:s3_posewise} shows the pose-wise median TDE improvement of Fused registration relative to Therapeutic only.

